\documentclass[aps,prl,reprint,showpacs,superscriptaddress]{revtex4-1}
\usepackage{amsfonts,mathrsfs,amsmath,amsthm,amssymb,bbold}
\usepackage{epsfig}
\usepackage{bm,mathrsfs}
\usepackage{lineno,hyperref}
\usepackage{amsmath}
\usepackage{color}
\usepackage{xcolor,colortbl} 
\usepackage{soul}
\usepackage{afterpage}
\usepackage{capt-of}
\usepackage{makecell}
\usepackage{graphicx}
\setcellgapes{4pt}

\newcolumntype{P}[1]{>{\centering\arraybackslash}p{#1}}

\usepackage{amsmath} 

\begin{document}
\sloppy
\clearpage

\title[mode = title]{Dzyaloshinskii--Moriya interaction in Nd$_{2}$Fe$_{14}$B as the origin of spin reorientation and rotating magnetocaloric effect}

%
\author{Hung Ba Tran}
\thanks{Electronic address: tran.h.ag@m.titech.ac.jp}
\address{Laboratory for Materials and Structures, Institute of Innovative Research, Tokyo Institute of Technology, Midori-ku, Yokohama 226-8503, Japan}
\address{Quemix Inc., Taiyo Life Nihombashi Building, 2-11-2, Nihombashi, Chuo-ku, Tokyo 103-0027, Japan}

\author{Yu-ichiro Matsushita}
\address{Laboratory for Materials and Structures, Institute of Innovative Research, Tokyo Institute of Technology, Midori-ku, Yokohama 226-8503, Japan}
\address{Quemix Inc., Taiyo Life Nihombashi Building, 2-11-2, Nihombashi, Chuo-ku, Tokyo 103-0027, Japan}
\address{Quantum Material and Applications Research Center, National Institutes for Quantum Science and Technology, 2-12-1, Ookayama, Meguro-ku, Tokyo 152-8552, Japan}
\date{\today}

\begin{abstract}
The mechanism of spin reorientation in Nd$_{2}$Fe$_{14}$B, which is a host crystal of a well-known neodymium permanent magnet, is studied by combining first--principles calculations and Monte Carlo simulations. The spin reorientation is thought to be derived from crystal field effects and gets less attention because of the undesirable property for hard magnet application. Dzyaloshinskii--Moriya interactions are usually less attractive or often ignored in rare--earth bulk systems, including permanent magnets such as Nd$_{2}$Fe$_{14}$B since people believe that the magnetic anisotropy is more dominant than the Dzyaloshinskii--Moriya interactions. However, in this study, we have found, for the first time, that the spin reorientation in Nd$_{2}$Fe$_{14}$B is attributed to Dzyaloshinskii--Moriya interactions. We have found, furthermore, the spin reorientation in Nd$_{2}$Fe$_{14}$B yields a great stage of rotating magnetocaloric effect at practical application level. We have found that the Dzyaloshinskii--Moriya interactions definitely contributes to the physical properties as a non--negligible effect in magnetic materials.
\end{abstract}

\maketitle

\begin{figure}[h!]
\centering
\includegraphics[width=8.6cm]{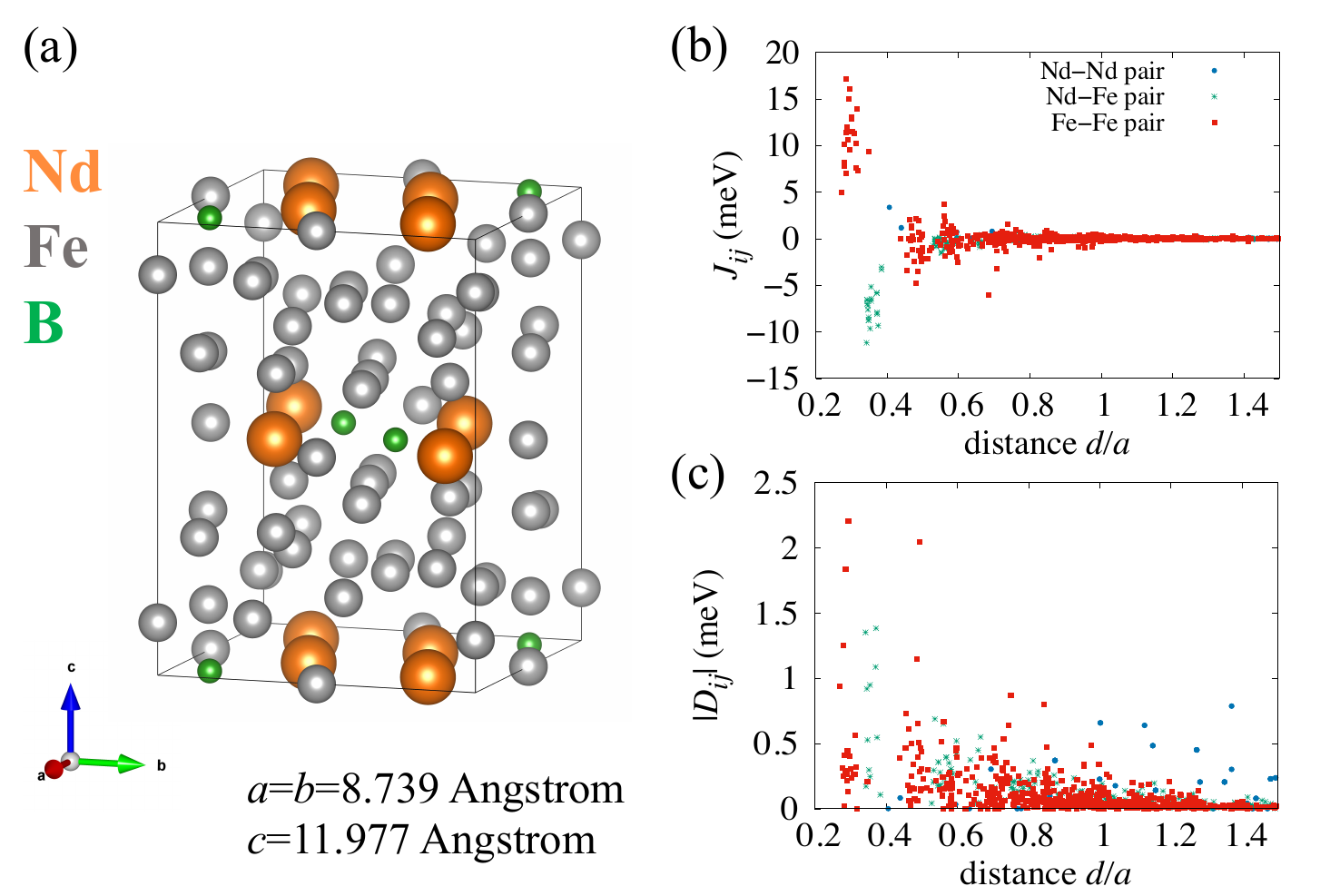} 
\caption{(a) Crystal structure of Nd$_{2}$Fe$_{14}$B where Nd, Fe, and B are indicated as orange, gray, and green spheres, respectively. The magnetic exchange coupling constants (b) and Dzyaloshinskii--Moriya vector lengths (c) of Nd--Nd (blue), Nd--Fe (green), and Fe--Fe (red) pairs. The color of Dzyaloshinskii--Moriya vector lengths for Nd--Nd, Nd--Fe, and Fe--Fe pairs is the same as the magnetic exchange coupling constants. } 
\label{FIG1}
\end{figure}

Sustainable energy brings many advantages to human life, such as power generation and electric vehicles\cite{Steven2012Nat}. Improving the efficiency of the permanent magnet is the core technology, which is the main purpose of many companies and governments\cite{Steven2012Nat,Steven2017NatMat}. Among many hard magnetic materials that have been discovered last four decades, neodymium permanent magnet, which has chemical formula Nd$_{2}$Fe$_{14}$B, remains as an unbeatable material\cite{David2002JMMM,Coey2020Eng}. The main reasons for this are the high Curie temperature, magnetocrystalline anisotropy energy, saturation magnetization, and coercivity of the material\cite{David2002JMMM,Coey2020Eng}. Crystal structure of Nd$_{2}$Fe$_{14}$B is shown as FIG. \ref{FIG1} (a), which has the space group $P$4$_{2}$/$mnm$ (\#136). However, the material also has a drawback at low temperatures, called the spin reorientation effect\cite{Durst1986JMMM,Hirosawa1987JAP}. The transition from uniaxial anisotropy to the cone magnetization at 135 K in the experiment leads to a significant decrease of magnetic anisotropy and magnetic flux density. Understanding the mechanism of these phenomena draws many advantages to improve the material's performance, which is interesting in both fundamental and industrial. The mechanism of spin reorientation in the material is studied by the phenomenological model, where the important part is the crystal field, which is usually taken from the experimental data\cite{Ito2016JMMM,Sasaki2015APEX,Toga2016PRB}. It limits the theoretical prediction to the Nd$_{2}$Fe$_{14}$B and also new material.

It motivates us to study the origin of spin reorientation of Nd$_{2}$Fe$_{14}$B by combining first-principles calculations and Monte Carlo simulations. In this work, we derive Hamiltonian from first-principles calculations, where the essential parts are isotropic exchange, antisymmetric exchange (also known as Dzyaloshinskii--Moriya interaction), magnetic anisotropy, and Zeeman term are taken into account. The Hamiltonian of Heisenberg model is used as 

\begin{equation}
\begin{split}
&H_{\rm Heis}=-\sum_{<ij>}J_{ij}^{m}\overrightarrow{S_{i}}\cdot\overrightarrow{S_{j}} -\sum_{i}k_{\rm u}(\overrightarrow{e_{u}}\cdot\overrightarrow{S_{i}})^{2} \\
& -\sum_{<ij>}\overrightarrow{D_{ij}}\cdot(\overrightarrow{S_{i}} \times \overrightarrow{S_{j}}) -g\mu _{\rm B}\sum_{i}\overrightarrow{H_{\rm ext}}\cdot\overrightarrow{S_{i}},
\end{split}
\label{Eq1}
\end{equation}  

\begin{figure}[h!] 
\centering
\includegraphics[width=8.6cm]{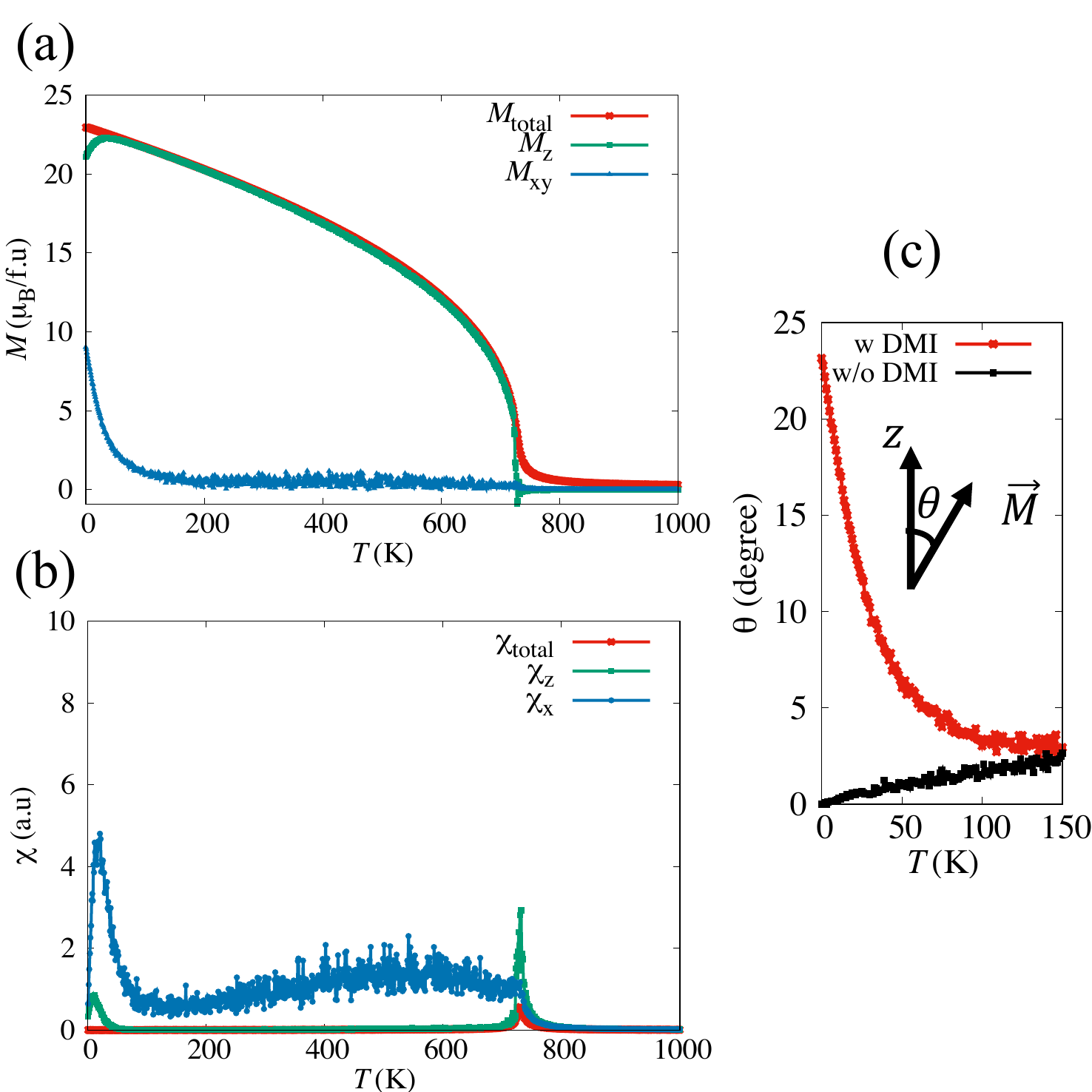} 
\caption{(a) Total (red) and projected of spontaneous magnetization on out--of--plane (green) and in--plane (blue) as a function of temperature. (b) Temperature dependence of magnetic susceptibility of total magnetization (red) and projected magnetization of $z$ direction (green) and $x$ direction (blue). (c) Temperature dependence of the tilt angle between magnetization direction and $z$ direction in the case of switch on Dzyaloshinskii--Moriya interaction (red) and switch off Dzyaloshinskii--Moriya interaction (black). } 
\label{FIG2}
\end{figure}

\noindent where $g$ is the $g$-factor, $\mu_{\rm B}$ is the Bohr magneton, $k_{\rm u}$ is the uniaxial anisotropy constant. The first term expresses the exchange interactions between spins at sites \textit{i} and \textit{j}. The spin tends to be parallel to the neighboring site when the magnetic exchange coupling constant $J_{ij}^{m}$ is positive. The spin is antiparallel to the neighboring spin for negative $J_{ij}^{m}$, which can be calculated by using the Liechtenstein formula in density functional theory (DFT)\cite{Liechtenstein1987JMMM,EbertSPRKKR,Ebert2011RPP}. According to our calculations of FIG. \ref{FIG1} (b), the positive coupling constant of the Fe--Fe pair is the main contribution for magnetism at high temperature, while the spin of Nd and Fe prefer antiparallel as ferrimagnetic  (FiM) order. In fact, the total energy of FiM is lower than that of ferromagnetic (FM) order by 0.059 (eV/atom) in DFT. The second term in Eq. (\ref{Eq1}) is the interaction between the spin at site \textit{i} and the uniaxial anisotropy, $\overrightarrow{e_{u}}$ being the direction of the easy axis in the case of a positive $k_{\rm u}$. $\overrightarrow{e_{u}}$ is set as out--of--plane ($c$ direction) and $k_{\rm u}$ is evaluated from the value of magnetocrystalline anisotropy energy (MAE) in first-principles calculations. Our calculated MAE is 9.58 (meV/f.u) equivalent to 6.72 (MJ/m$^{3}$) with out-of-plane easy axis\cite{Kresse1996PRB}. It is in good agreement with the experiment value 4.9 (MJ/m$^{3}$) and also direction at 300 K\cite{Durst1986JMMM,Hirosawa1987JAP}. The third term is the antisymmetric exchange term with $\overrightarrow{D_{ij}}$ being the Dzyaloshinskii--Moriya vector, which can be calculated as the slope of magnon at $\Gamma$ point in the case of spin spiral calculations\cite{Mankovsky2017PRB}. So far, the Dzyaloshinskii--Moriya interaction is usually treated as negligible for the Nd$_{2}$Fe$_{14}$B permanent magnet. The length of the Dzyaloshinskii--Moriya vector is shown as FIG. \ref{FIG1} (c). The length of the Dzyaloshinskii--Moriya vector has long-range behavior and comparable with magnetic anisotropy energy. The magnetic order of Nd$_{2}$Fe$_{14}$B can be cone magnetization below spin reorientation temperature due to the competition of Dzyaloshinskii--Moriya interaction and magnetic anisotropy. The final term is Zeeman, which is the effect of the external magnetic field on the spin.

As the results of Monte Carlo simulations, the calculated total and projected magnetization and magnetic susceptibility are shown in FIG. \ref{FIG2} (a), (b). At low temperature, the magnetization component along $z$ direction is slightly smaller than total magnetization since the in-plane magnetization is finite. The magnetization direction becomes parallel to the easy axis when the temperature is close to spin reorientation one, consistent with the transition from cone magnetization to uniaxial as experimental works\cite{Durst1986JMMM,Hirosawa1987JAP}. The total magnetic susceptibility, which is the fluctuation of total magnetization, only has one peak at 728 K (the Curie temperature), while the $z$ component shows two clear peaks corresponding to the spin reorientation and ferromagnetic--paramagnetic (FM--PM) transition. On the other hand, the $x$ component of magnetic susceptibility only diverges at the transition temperature between cone magnetization and uniaxial. The projected magnetization in Monte Carlo simulation shows that the tilted angle between magnetization and easy axis can be calculated as FIG. \ref{FIG2} (c). The value of the tilted angle is quantitatively in agreement with the previous works\cite{Toga2016PRB}. The tilt angle decreases when temperature increases due to the weaken of Dzyaloshinskii--Moriya interaction. It differs from the Nd atoms' crystal field term, which is proposed as the origin of spin reorientation in previous works\cite{Ito2016JMMM,Sasaki2015APEX,Toga2016PRB}. We also consider the case of without Dzyaloshinskii--Moriya interaction. In this case, the tilt angle increases as temperature increases due to the thermal fluctuation.

\begin{figure}[h!] 
\centering
\includegraphics[width=8.6cm]{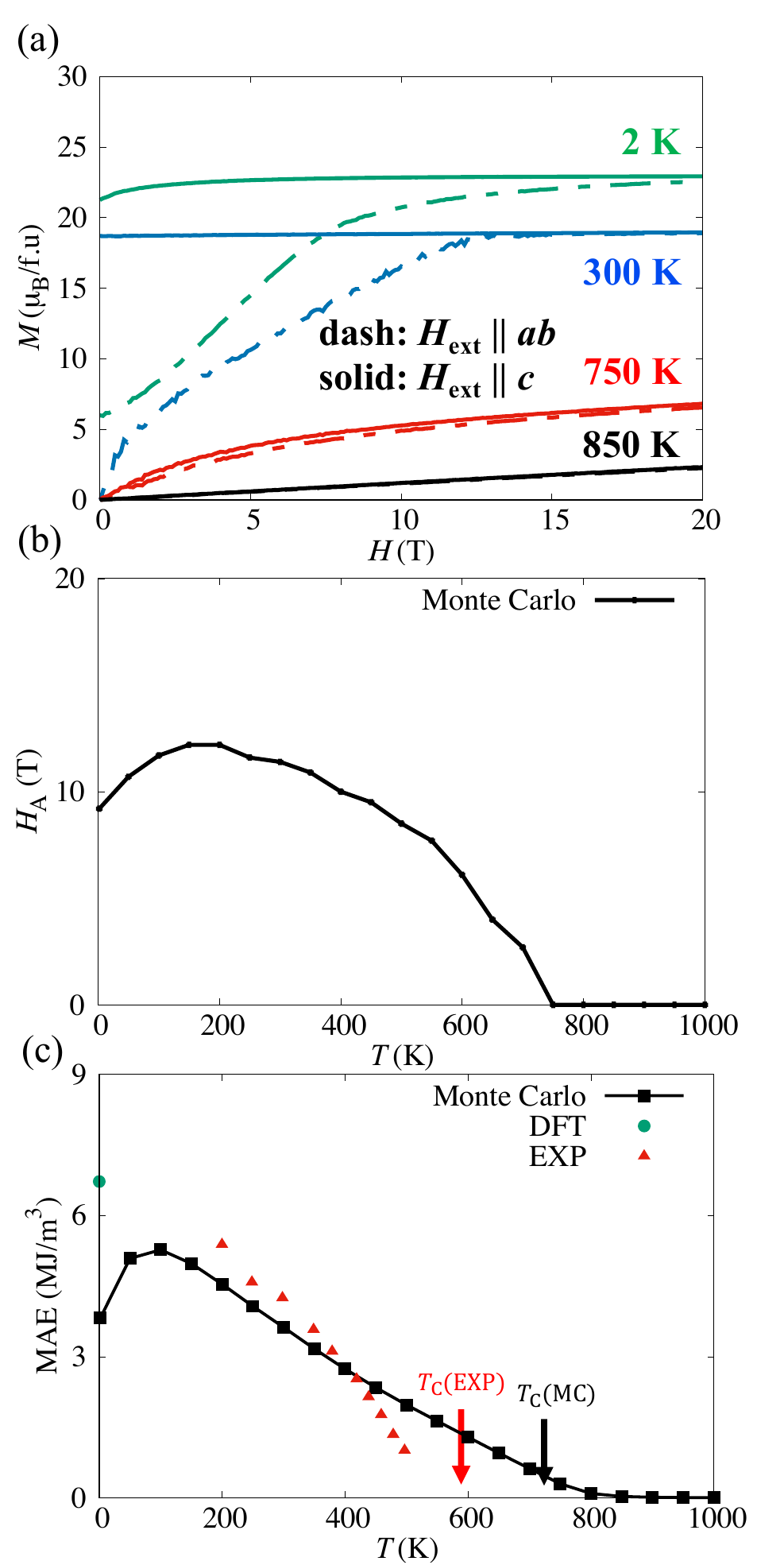} 
\caption{(a) $M$--$H$ curves when applying magnetic field along $ab$ plane (dash line) and easy axis (solid line) at low temperature (green), intermediate temperature (blue), slightly above Curie temperature (red), high temperature (black). (b) Temperature dependence of the anisotropy field by extrapolating the $M$--$H$ curves. (c) The magnetic anisotropy energy of DFT (green) and Monte Carlo method by integrating the $M$--$H$ curves (black) and experiment (red)\cite{Hirosawa1987JAP}. } 
\label{FIG3}
\end{figure}

The magnetization--magnetic field ($M$--$H$) at several temperatures is considered as FIG. \ref{FIG3} (a) to understand the effect of antisymmetric exchange term in magnetic anisotropy. At 2 K (below the spin reorientation), the magnetization has some component along the $ab$ plane at zero external magnetic fields. It assists the material to magnetize along the hard plane when applying a magnetic field along the $ab$ plane. Applying a magnetic field along the hard plane at 2 K has the slope higher than at 300 K (above the spin reorientation temperature) due to the reduction in the strength Dzyaloshinskii--Moriya interaction. The in--plane component of magnetization at above spin reorientation temperature becomes zero, it means that the direction of magnetization is along the out--of--plane. The $M$--$H$ curves show some area at 750 K slightly higher than the Curie temperature due to the effect of anisotropic magnetic susceptibility, which is studied in our previous works\cite{HBT2022PRB,HBT2022AM}. Note that the $M$--$H$ area is finite at this temperature while the magnetization at zero external magnetic fields is approximately equal to zero for both the easy axis and hard plane. The anisotropic magnetic susceptibility almost disappears at sufficiently high temperature. 

The temperature dependence of the anisotropy field, the magnetic field required to saturate the magnetization along the hard axis, can be evaluated by extrapolating the $M$--$H$ curves as FIG. \ref{FIG3} (b). The anisotropy field has a peak at spin reorientation temperature as the effect of the Dzyaloshinskii--Moriya interaction. After getting the maximum, the anisotropy field decreases as temperature increases and becomes zero at the Curie temperature. The anisotropy field can be roughly estimated as $H_{\rm A}=2K_{\rm A}/M_{\rm s}$ by using Stoner--Wohlfarth model for uniaxial anisotropy without Dzyaloshinskii--Moriya interaction. In this case, the anisotropy field always decreases as temperature increases since the effective anisotropy constant decreases much faster than magnetization in Callen--Callen theory. Due to the effect of the Dzyaloshinskii--Moriya interaction at low temperature, the anisotropy field increases as temperature increases, which is in good agreement with experimental studies\cite{Hirosawa1987JAP}. Note that the location of the peak in the anisotropy field is higher than the peak of magnetic susceptibility of the $x$ component. The anisotropy field is highest when the tilted angle is smallest or the minimum of magnetic susceptibility of $x$ axis, as the disappearance of Dzyaloshinskii--Moriya interaction.

The magnetic anisotropy energy of Monte Carlo, DFT, and experimental work are shown in FIG. \ref{FIG3} (c). Due to the effect of Dzyaloshinskii--Moriya interaction on $M$--$H$ curves, the magnetic anisotropy energy, which is estimated by integrating the $M$--$H$ curves, is underestimated compared with the value of DFT at low-temperature region. The peak location is similar to the peak of the anisotropy field. At the above spin reorientation temperature, the magnetic anisotropy energy decreases as temperature increases and becomes negligible at a slightly higher temperature than the anisotropy field. The anisotropy field becomes approximately zero if the spontaneous magnetization is negligible. On the other hand, the magnetic anisotropy energy is still finite at slightly above Curie temperature due to anisotropic magnetic susceptibility. Our result is semiquantitatively in agreement with experimental work\cite{Hirosawa1987JAP}. These fact are the fingerprint of the important role of the Dzyaloshinskii--Moriya interaction.

\begin{figure}[h!] 
\centering
\includegraphics[width=8.6cm]{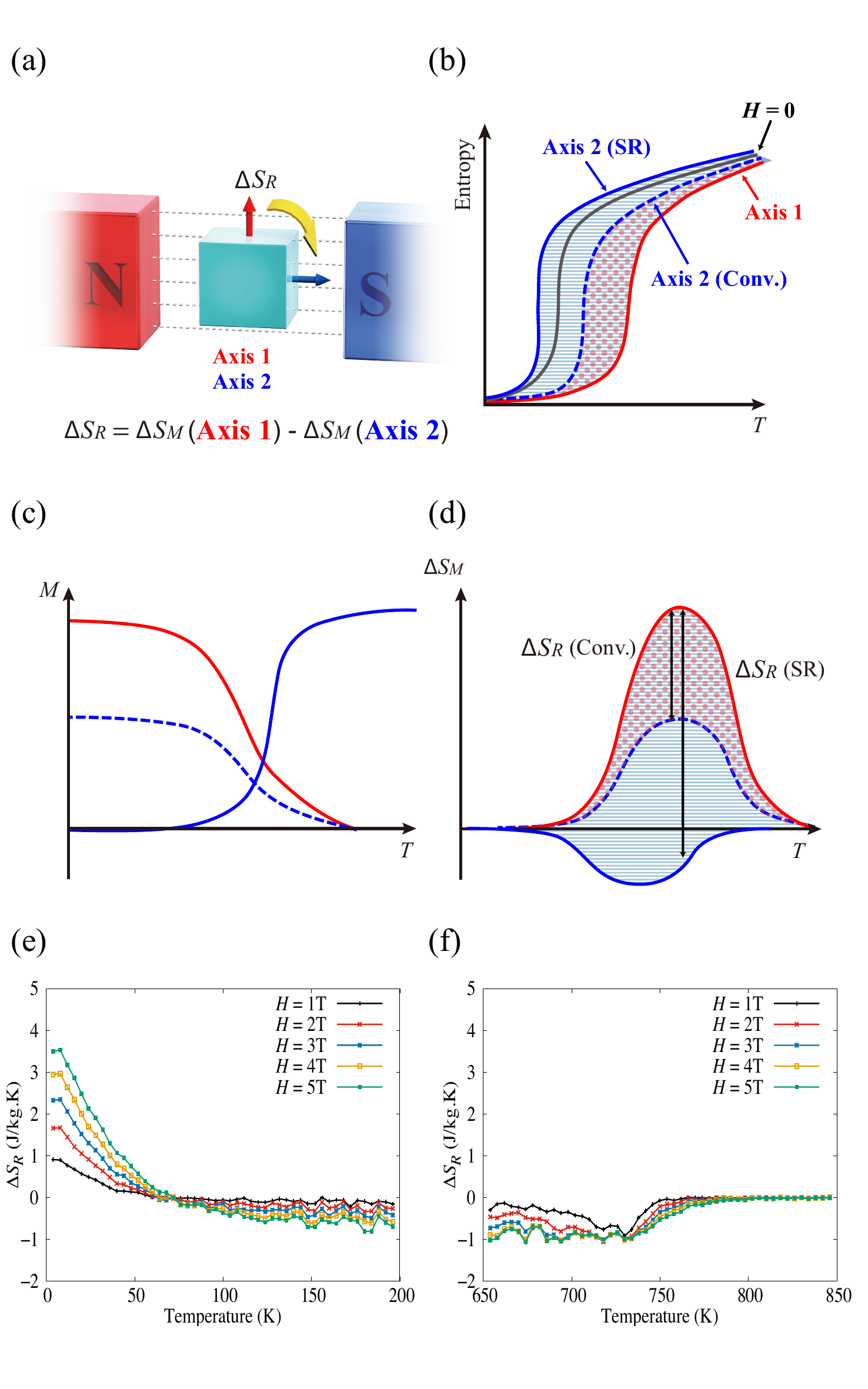} 
\caption{The diagram of rotating magnetocaloric effect (a), where $\Delta S_{R}$ and ($\Delta S_{M}$) are rotating magnetic entropy change and isothermal magnetic entropy change. The diagram of rotating magnetocaloric effect of usual (conventional) and unusual (spin reorientation) of total entropy change curves (b), magnetization curves (c), and isothermal magnetic entropy change (d). Rotating magnetic entropy change of Nd$_{2}$Fe$_{14}$B as the function of temperature and magnetic field around spin reorientation temperature (e) and Curie temperature (f). } 
\label{FIG4}
\end{figure}

We have clarified the origins of the spin reorientation as Dzyaloshinskii--Moriya interaction. We have also found that the spin reorientation, which has not got attention so much so far, provides an ideal stage for rotating magnetocaloric effect as showed in FIG. \ref{FIG4} (a), (b), (c), and (d). Rotating magnetocaloric effect relates to controlling the heat absorb or release by changing the relative direction of the external magnetic field with crystal structure, rotating the sample at fixed strength of the external magnetic field(See FIG. \ref{FIG4} (a)). The conventional rotating magnetocaloric effect mainly comes from the modification of magnetocrystalline anisotropy at ferromagnetic--paramagnetic (FM--PM) transition(See FIG. \ref{FIG4} (c)). The isothermal magnetic entropy change of easy and hard axis have the same sign since the magnetic field applied leads to higher critical temperature(See FIG. \ref{FIG4} (b) and (d))). The rotating magnetic entropy change, which is the difference of isothermal magnetic entropy change, should be smaller than the amplitude of isothermal magnetic entropy change in this case. On the other hand, in the spin reorientation effect, applying the external magnetic field on different axis leads to opposite effect in magnetization since the external magnetic field tends to expand the region of projected magnetization on the direction of external magnetic field(See FIG. \ref{FIG4} (d))). It is promising as a possible way to enhance the rotating magnetic entropy change to higher than the isothermal magnetic entropy change since the entropy change of easy and hard axis has different signs.

The isothermal magnetic entropy changes ($\Delta S_{M}$) are estimated by integrating the derivative of the magnetization ($M$) based on the Maxwell relation\cite{HBT2022PRB,HBT2022AM}

\begin{equation}
\begin{split}
& \Delta S_{M}(H_{\rm ext},T) =\int_{0}^{H_{\rm ext}}\left (\frac{\partial M(H,T) }{\partial T} \right )dH, \\
& \cong \sum_{j=0}^{N}\frac{M(H_j,T+\Delta T)-M(H_j,T-\Delta T)}{2\Delta T}\Delta H
\end{split}
\label{Eq2}
\end{equation}

\noindent The isothermal magnetic entropy change is obtained by integrating with fine mesh in temperature and external magnetic field. Due to the magnetocrystalline anisotropy energy, the isothermal magnetic entropy change strongly depends on the direction of the applied magnetic field. The rotating magnetic entropy can be calculated as the difference in isothermal magnetic entropy change when applying magnetic fields along the easy and hard axes. We assume that rotating the sample at zero external magnetic field causes no heat. The rotating magnetic entropy ($\Delta S_{R}$) relates to the amount of heat can be used for cooling by rotating the sample at fix strength of magnetic field, which can be estimated as the difference of $\Delta S_{M}$ of two directions.

The rotating magnetic entropy, which is related to the amount of heat that can be used for cooling by rotating the sample from the hard axis to the easy axis at the fixed strength of the external magnetic field, is shown in FIG. \ref{FIG4} (e), (f). The usual rotating magnetic entropy relates to the effect of magnetic anisotropy energy at FM--PM transition, which is limited to the absolute value of isothermal magnetic entropy change since the easy axis and hard axis have the same sign. On the other hand, the unusual rotating magnetic entropy of spin reorientation transition can be higher than the absolute value of isothermal magnetic entropy change since the sign of them can be opposite. The isothermal magnetic entropy change of the easy axis at low temperature has a positive sign since it is estimated as the integration of the magnetization curve along the $z$ direction. On the other hand, the isothermal magnetic entropy change of the hard axis has a negative sign as usual FM--PM transition since the in--plane magnetization decreases when temperature increases. It enhances the unusual rotating magnetic entropy in the case of spin reorientation, which is almost four times larger than the usual case of FM--PM transition. The location of the entropy peak of low temperature is similar to the peak of in--plane magnetic susceptibility, while the entropy curve crossed at zero at spin reorientation temperature.

In summary, we have found that the origin of spin reorientation of well--known Nd$_{2}$Fe$_{14}$B permanent magnet comes from the effect of Dzyaloshinskii--Moriya interaction. The result of magnetization curves and tilted angle is well reproduced of the previous work using crystal electric field interaction. The temperature dependence of $M$--$H$ curves, anisotropy field, and magnetic anisotropy energy are in good agreement with experimental works. Due to the effect of the Dzyaloshinskii--Moriya interaction, magnetizing the material along with the in--plane direction at low temperature becomes easier than above spin reorientation temperature. The rotating magnetic entropy change can be significantly enhanced around spin reorientation transition. The peak of spin reorientation of 5 T is almost four times higher than those values of the usual ferromagnetic--paramagnetic transition.

\section*{Methods}
The relaxed crystal structure parameters and magnetocrystalline anisotropy energy of Nd$_{2}$Fe$_{14}$B are obtained by using the VASP code. The magnetic exchange coupling constants and Dzyaloshinskii--Moriya vectors are calculated based on the Green function multiple-scattering formalism as implemented in SPR--KKR code. The Monte Carlo simulations are performed by using the in--house program.

\section*{Data availability}
The datasets generated and/or analyzed during the current study are available from the corresponding author on reasonable request.

\section*{Acknowledgements}
This work was supported by MEXT as ”Program for Promoting Researches on the Supercomputer Fugaku” (JP-MXP1020200205) and JSPS KAKENHI as ”Grant-in-Aid for Scientific Research(A)” Grant Number 21H04553. The computation in this work has been done using supercomputer Fugaku provided by the RIKEN Center for Computational Science, Supercomputer Center at the Institute for Solid State Physics in the University of Tokyo, and the TSUBAME3.0 supercomputer at Tokyo Institute of Technology.

\section*{Author Contributions}
H. B. T. invented the method and wrote the paper with the supervision of Y. M., all authors have given approval to the final version of the manuscript.

\section*{Competing interests}
The authors declare no competing interests.

\bibliography{basename of .bib file}

\end{document}